\documentclass[conference]{IEEEtran}
\IEEEoverridecommandlockouts
\usepackage{cite}
\usepackage{amsmath,amssymb,amsfonts}
\usepackage{algorithmic}
\usepackage{graphicx}
\usepackage{textcomp}
\usepackage{xcolor}
\usepackage{booktabs}   
\usepackage{pifont}     
\usepackage{subcaption} 
\usepackage{caption} 
\usepackage{booktabs}
\usepackage{enumitem} 
\usepackage{multirow}
\usepackage{hyperref} 
\usepackage{url}
\usepackage{float}
\def\BibTeX{{\rm B\kern-.05em{\sc i\kern-.025em b}\kern-.08em
    T\kern-.1667em\lower.7ex\hbox{E}\kern-.125emX}}

\makeatletter
\newcommand{\linebreakand}{%
  \end{@IEEEauthorhalign}
  \hfill\mbox{}\par
  \mbox{}\hfill\begin{@IEEEauthorhalign}
}
\makeatother
\begin{document}

\title{Unsupervised Backdoor Detection and Mitigation for Spiking Neural Networks}



\author{\IEEEauthorblockN{Jiachen Li}
\IEEEauthorblockA{\textit{RMIT University} \\
Melbourne, Australia \\
jiachen.li@rmit.edu.au}
\and
\IEEEauthorblockN{Bang Wu}
\IEEEauthorblockA{\textit{RMIT University} \\
Melbourne, Australia \\
bang.wu@rmit.edu.au}
\and
\IEEEauthorblockN{Xiaoyu Xia}
\IEEEauthorblockA{\textit{RMIT University} \\
Melbourne, Australia \\
xiaoyu.xia@rmit.edu.au}
\linebreakand 
\IEEEauthorblockN{Xiaoning Liu}
\IEEEauthorblockA{\textit{RMIT University} \\
Melbourne, Australia \\
xiaoning.liu@rmit.edu.au}
\and
\IEEEauthorblockN{Xun Yi}
\IEEEauthorblockA{\textit{RMIT University} \\
Melbourne, Australia \\
xun.yi@rmit.edu.au}
\and
\IEEEauthorblockN{Xiuzhen Zhang}
\IEEEauthorblockA{\textit{RMIT University} \\
Melbourne, Australia \\
xiuzhen.zhang@rmit.edu.au}
}
\maketitle

\begin{abstract}
Spiking Neural Networks (SNNs) have attracted significant attention from the research community due to their high energy efficiency compared to Artificial Neural Networks (ANNs). However, rare studies on the security of SNNs were conducted, especially in backdoor attacks. Existing defense methods for ANN backdoor attacks either perform poorly or can be easily bypassed in SNN scenarios due to SNNs' event-driven and temporal dependency characteristics, posing significant research challenges. In this paper, we identify the blockers to existing backdoor defenses for defending against attacks in SNNs and propose an unsupervised post-training backdoor detection method named Temporal Membrane Potential Backdoor Detection (TMPBD) to address those blockers in SNNs with neuromorphic data. Specifically, TMPBD employs the maximum margin statistic of temporal membrane potential in the last spiking layer of the SNNs to detect attack target labels without knowledge of the attack or access to any data.
Moreover, we also design a practical and robust mitigation mechanism named Neural Dendrites Suppression Backdoor Mitigation (NDSBM). NDSBM dually clamps the neural dendrites, i.e., the weights connecting the first two convolution layers in each convolution block to limit the backdoor effect, while preserving the benign model behaviors learned from the temporal membrane potential obtained from a small, clean, unlabeled dataset in the same domain. To evaluate the performance, we conduct a comprehensive evaluation with multiple backdoor attack techniques, including the SOTA input-aware dynamic trigger attack dedicated to SNNs with clean models on three neuromorphic benchmark datasets. The results demonstrated that TMPBD achieves 100\% prediction accuracy in detecting dynamic trigger attacks and associating attack target labels in all benchmark datasets. NDSBM lowered the attack success rate (ASR) from 100\% caused by the dynamic trigger attack down to 8.44\% with only mitigation or 2.81\% when combined with detection for an end-to-end pipeline without performance degradation in clean accuracy. 

\end{abstract}


\section{Introduction}

Spiking Neural Networks (SNNs) \cite{ghosh-dastidar_spiking_2009,tavanaei_deep_2019, eshraghian_training_2023,fang_incorporating_2021}, inspired by the biological neural processes of the human brain \cite{gerstner_neuronal_2014,eshraghian_training_2023}, are a promising alternative to Artificial Neural Networks (ANNs) due to their spatio-temporal, discrete representation, and event-driven properties that significantly reduce power consumption \cite{kim_spiking-yolo_2020, gerstner_spiking_2002}. A recent study \cite{kundu_spike-thrift_2021} suggests that SNNs can achieve 12.2 times energy efficiency compared to ANNs with a similar number of parameters. Performance was once the weakness of SNNs but not anymore under recent technological leaps, with major milestones achieving performance comparable to ANNs \cite{tavanaei_deep_2019} in complex tasks such as autonomous driving \cite{chen_event-based_2020,viale_carsnn_2021}, computer vision \cite{escobar_action_2009}, speech recognition \cite{loiselle_exploration_2005} and medical diagnosis \cite{kasabov_evolving_2014}. SNNs were naturally designed to work with neuromorphic data captured by Dynamic Vision Sensor (DVS) cameras \cite{serrano-gotarredona_128_2013}. Unlike traditional cameras, which capture the absolute brightness of RGB lights at a constant frame rate, the DVS camera captures independent discrete events that describe the change in light intensity at certain pixels. The event-driven and sparse nature of events enables the neuromorphic data to improve energy efficiency and temporal resolution while minimizing latency.

Despite the advantages, SNNs remain vulnerable to a range of security threats faced by ANNs~\cite{li_backdoor_2022,wang_siguard_2025}, notably insidious backdoor attacks~\cite{ abad_sneaky_2024}. In a backdoor attack, an adversary injects a hidden trigger during training so that the model produces an attacker–specified output whenever the trigger is present, while behaving normally on trigger-free inputs~\cite{gu_badnets_2019}. Such covert manipulation can undermine model integrity in mission-critical applications, e.g., allowing an attacker to bypass the facial-recognition security checkpoint~\cite{li_light_2020}.


Research in backdoor attacks in SNNs has made great progress, with the state-of-the-art (SOTA) dynamic trigger backdoor attack designed for neuromorphic data in SNNs achieving 100\% attack success rate (ASR) with negligible degradation in clean accuracy (CA) and undetectable to human inspection \cite{abad_sneaky_2024}. However, to our best knowledge, the research on backdoor defense in SNNs is remarkably scarce, where there are no dedicated backdoor defense frameworks proposed for SNNs. The existing defenses adopted from ANN to SNN are poorly performed or can be easily passed by adaptive attacks \cite{abad_poster_2022,abad_sneaky_2024,riano_flashy_2024}. The main challenge comes from the fundamental difference in neuron behavior and data format from neuromorphic data in SNNs to static images in ANN, requiring a complete redesign of the backdoor defense algorithm, specifically accommodating the characteristics of SNNs.

In this paper, we first investigate the deficiencies of existing ANN backdoor defenses. Then we propose the first full-lifecycle backdoor detection and mitigation framework dedicated to SNNs in a strictly practical, data-free setting. The proposed Temporal Membrane Potential Backdoor Detection (TMPBD) innovatively uses temporal membrane potential (TMP) and maximum margin (MM) statistics-based anomaly detection to detect the backdoor attack target label. The proposed Neural Dendrites Suppression Backdoor Mitigation (NDSBM) uses clamping-based mitigation to reduce the backdoor effect.

To ensure robustness and practical relevance, we conducted a comprehensive experiment on proposed frameworks and discussed potential threats to validity. 

In summary, our contribution includes: 
\begin{itemize}[topsep=1pt]
    \item We adopt several renowned backdoor defense strategies in ANNs to SNNs and analyze the challenges blocking them from being as effective in SNNs. Based on those findings, we propose innovative designs to solve the identified challenges to defending against backdoor attacks in SNNs with neuromorphic data. 
    \item We propose TMPBD, a novel data-free, unsupervised backdoor detection strategy based on the TMP's MM statistic, which reaches 100\% attack label detection accuracy on models poisoned by various backdoor attacks without access to any data. 
    \item We propose NDSBM, a novel unsupervised backdoor mitigation strategy based on clamping the weights of the connection, also known as neural dendrites in SNNs, between the first two convolution layers in each convolution block of the model. NDSBM is capable of lowering the ASR from 100\% down to 8.44\% on average against dynamic trigger attacks. In addition, we utilize the end-to-end backdoor defense pipeline for both proposed backdoor detection and mitigation strategies to further reduce the ASR under SOTA dynamic trigger attack to 2.81\% on average while achieving higher CA. 
    \item We comprehensively evaluate the proposed backdoor defense strategies against the existing defense methods adopted for ten repetitions with multiple attack types and variant datasets. 
    \item We critically discuss the scalability and robustness of the proposed methods against imbalanced datasets and adaptive attackers and provide indicative solutions to false-positive, intrinsic backdoor, and all-to-all attack issues when additional information are available. 
\end{itemize}

\section{Preliminaries}
This section introduces the essential terminologies, concepts, and notations of SNNs and backdoor attacks to supplement the preliminary knowledge needed for the subsequent paper sections. 
\subsection{Spiking Neural Network}
\label{snn}

The SNNs are described as the third generation of neural network machine learning models known for improved energy efficiency over their predecessor, ANNs \cite{maass_networks_1997}. As a class of deep neural networks, SNNs inherit the same network structures from fully connected ANNs, with interconnected input, hidden, and output layers. Inspired by biological neurons \cite{kim_spiking-yolo_2020, gerstner_spiking_2002}, the neurons in SNNs emit discrete spike events to pass the information enclosed in spike timing \cite{tavanaei_deep_2019}. The neurons emit spikes only when the accumulated input current exceeds the threshold \cite{bu_rate_2023}. In contrast, ANNs transmit information in continuous-valued signals and employ activation functions to capture non-linear relationships \cite{fukushima_cognitron_1975}. To simulate SNNs on modern computers with the Von Neumann architecture, it is a common practice to simplify the operation by discretizing the time. Where the behavior of spiking neurons following the representative Leaky-Integrate-and-Fire (LIF) model can be described mathematically \cite{fang_incorporating_2021} as follows :

\begin{equation} \label{eq:1}
H_t = V_{t - 1} + \frac{1}{\tau} \left( X_t - \left( V_{t - 1} - V_{\text{reset}} \right) \right)
\end{equation}
\begin{equation} \label{eq:2}
S_t = \Theta(H_t - V_{\text{threshold}})
\end{equation}
\begin{equation} \label{eq:3}
\Theta(x) = 
\begin{cases}
1, & x \geq 0 \\
0, & x < 0
\end{cases}
\end{equation}
\begin{equation} \label{eq:4}
V_t = H_t \cdot (1 - S_t) + V_{\text{reset}} \cdot S_t
\end{equation}

Equation~\eqref{eq:1} describes the dynamics of a leaky integrate-and-fire (LIF) neuron: the membrane potential \(V_t\), an internal state that integrates the input and leaks over time. Here, \(H_t\) denotes the instantaneous (pre-spike) membrane potential after integration/charging and before firing, \(X_t\) is the input at time \(t\), and \(V_{t-1}\) is the post-spike membrane potential from the previous time step. The membrane time constant \(\tau\) governs the decay of the potential toward the reset value \(V_{\text{reset}}\). The parameters \(V_{\text{reset}}\) and \(V_{\text{threshold}}\) are fixed properties of the neuron. Equations~\eqref{eq:2}--\eqref{eq:3} specify the spike-generation and reset rules: the neuron emits a spike \((S_t=1)\) if and only if \(H_t \geq V_{\text{threshold}}\); upon spiking, the membrane potential is reset to \(V_{\text{reset}}\).

In the input layer, \( X_t \) denotes the input from the neuromorphic data captured by DVS cameras \cite{serrano-gotarredona_128_2013}.  A DVS camera is different from a regular camera in that it captures absolute RGB brightness at a constant rate for all pixels. The DVS camera captures a series of events asynchronously. The event contains information on per-pixel brightness changes. An individual event can be described by set $(t, x, y, p)$, which denotes the event's timing, the pixel's x-y coordinate, and polarity. The polarity indicates the direction of change in brightness, lighter or darker. 

In hidden layers, the input \( X_t = \sum_{j} W_{ij} S_{j,t}\) is the weighted sum of outputs from nodes in the previous layer. The \( W_{ij}\) are the learnable weights representing the strength and direction of the connection from neuron \(j\) in the previous layer to \(i\). In the training stage, surrogate gradients \cite{neftci_surrogate_2019} that approximate a derivative of Equation \eqref{eq:3} enable backpropagation training on SNNs with Adam \cite{kingma_adam_2014} or stochastic gradient descent \cite{amari_backpropagation_1993} where the latter one is more popular for better performance \cite{lee_training_2016} thus utilized in this research.

In the output layer, the output of SNNs, the firing rate (FR), equivalent to logits in ANNs, is represented as \(FR = \frac{1}{T} \sum_{t=1}^T S_t\). Taking the Softmax of the FRs provides the label probability distribution for classification problems.

\subsection{Backdoor Attack}
\label{backdoor_attack}
The Backdoor Attack is one of the major security threats to machine learning models. A malicious attacker embeds a hidden trigger into a model during training, causing the model to misclassify specific inputs at inference time when the trigger is present. 

In a general ML pipeline, the classifier $h(x, \mathcal{D}) = \arg\max_{y} p(y | x, \mathcal{D})$ is trained to infer the most probable label based on the input sample $x$, and the training set $\mathcal{D} = \left\{ (x_1, y_1), \dots, (x_n, y_n) \right\}$ \cite{weber_rab_2023}. During a data poisoning-based dirty label all-to-one backdoor attack, the adversary manipulates the training data by adding a set of trigger patterns from the set $\Omega_x$ to a subset of samples and incorrectly labeling them as the attack target label $\tilde{y}$. Each sample-specific trigger pattern $\delta_i$ from the trigger pattern set $\Omega_x$ can be a pixel pattern, a color patch, or a specific shape. The poisoned data set with $r$  poisoned samples out of  $n$ total samples is denoted below:
\begin{equation}
\label{eq:poison_data}
  \mathcal{D}_{BD}(\Omega_x, \tilde{y}) = \left\{ (x_i + \delta_i, \tilde{y}_i) \right\}_{i=1}^{r} \cup \left\{ (x_i, y_i) \right\}_{i=r+1}^{n}
\end{equation}

As a result, the classifier trained on this poisoned dataset is compromised and denoted as: $h(x + \Omega_x, \mathcal{D}_{BD}(\Omega_x)) = \tilde{y}$. At inference time, the compromised classifier will misclassify the input as the nominated attack target class (ATC) $\tilde{y}$ when the trigger is presented. The attacker would carefully craft the trigger pattern $\Omega_x$ and decide on a poison rate $\frac{r}{n}$ that maximizes the performance of the attack, evaluated by the ASR denoted below:
\begin{equation}
\label{eq:asr}
    \max_{\Omega_x,r} \text{ASR} = \frac{\sum_{i=1}^{r} \mathbf{1} \left[ h(x_i + \delta_i, \mathcal{D}_{BD}(\Omega_x)) = \tilde{y} \right]}{r}
\end{equation}

In addition to maximizing the ASR, the attack is also motivated to maintain the model prediction accuracy on its originally designed task to ensure the poisoned model is being deployed by the victim smoothly without arousing suspicion. The performance of the model on the original task is evaluated by CA, as indicated below:
\begin{equation}
\label{eq:ca}
    \max_{\Omega_x,r} \text{CA} = \frac{\sum_{i=r+1}^{n} \mathbf{1} \left[ h(x_i, \mathcal{D}_{BD}(\Omega_x)) = y_i \right]}{n - r}
\end{equation}

To avoid attack trigger patterns being detected by pattern recognition defense algorithms or human inspection \cite{chen_badnl_2021}, the trigger patterns are usually motivated to minimize the L2 norm to ensure the stealthiness of the attack as denoted below:
\begin{equation}
\label{eq:stealthy}
 \min_{\Omega_x} \|\Omega_x\|_p \quad \text{s.t. } h(x + \Omega_x, \mathcal{D}_{BD}(\Omega_x)) = \tilde{y}, \, \forall x \in \mathcal{D}_{\text{clean}}
\end{equation}

The evaluation metric on stealthiness varies depending on the data format in the different problem domains, where the mean square error (MSE) \cite{wang_mean_2009} between the original and poison samples is commonly employed in the image domain. The structural similarity index metric (SSIM) is popular among neuromorphic data \cite{abad_sneaky_2024}.

Although the existing literature on backdoor attacks in SNNs is less than that of ANNs, existing research has successfully adopted several backdoor techniques from ANNs to SNNs with modification and achieved excellent performance. The consistency of effectiveness between SNNs and ANNs is because backdoor attacks mainly exploit the training process, where SNNs train similarly to ANNs with surrogate gradients \cite{neftci_surrogate_2019}. One of the adopted backdoor attacks on SNNs is the static trigger attack proposed in \cite{abad_poster_2022} inspired by the classic BadNet \cite{gu_badnets_2019}, aiming to conduct a content-independent fixed backdoor pattern attack with the poisoned dataset denoted as:
\begin{multline}
\label{eq:static_attack}
    \mathcal{D}_{BD}(\Omega_x, \tilde{y}) = \\ \left\{ (x_i^t +\delta_i, \tilde{y}) \right\}_{i=1, t=0}^{r, T} \cup \left\{ (x_i^t, y_i) \right\}_{i=r+1, t=0}^{n, T}
\end{multline}
The backdoor pattern $\Omega_x$ is constant in size, position, and polarity across all poisoned input $x_i$ in all time frames $t$ and often replaces the original value in the patched pixels. Attackers make a trade-off between a bigger patch size for higher ASR but lower stealthiness, or vice versa.

The current SOTA attack is the dynamic backdoor attack \cite{abad_sneaky_2024} inspired by the input-aware attack in ANNs \cite{nguyen_input-aware_2020,doan_lira_2021}. The dynamic attack is specifically designed for SNNs with machine-generated trigger patterns $\delta_i^t $ from $\Omega_x^t(x_i)$ that customize the sizes, positions, polarities, and distributions of the overlay pattern uniquely for each input sample $x_i$ at each time frame $t$. The dynamic trigger pattern is designed to bypass human inspection and pattern recognition-based machine detection during the inference stage. The compromised dataset with the dynamic trigger is denoted below:
\begin{multline}
\label{eq:dynamic}
    \mathcal{D}_{BD}(\Omega_x^t(x_i), \tilde{y}) = \\ \left\{ (x_i^t + \delta_i^t, \tilde{y}) \right\}_{i=1, t=0}^{r, T} \cup \left\{ (x_i^t, y_i) \right\}_{i=r+1, t=0}^{n, T}
\end{multline}
For the original paper by Abad et al. \cite{abad_sneaky_2024}, the authors demonstrated that their proposed dynamic trigger pattern backdoor attack achieved up to 100\% ASR, with negligible degradation in CA. In addition, the excellent stealthiness of the dynamic trigger patterns bypassed all human detection and posed a tremendous threat to the security of the SNN models.

\section{Problem Formulation}
This section introduces the setting and scenarios of the security risk that the paper proposes to defend.

\subsection{System Model}
In this paper, we consider a classical pre-trained model adoption scenario. We focus only on SNN models that take neuromorphic data as input and perform classic and common classification tasks. 

\subsubsection{Model provider} 
The model provider independently develops the SNN model and shares trained models with the model consumer.
\begin{itemize}[topsep=1pt]
    \item The model provider fully controls training data, including knowledge and modification capability.
    \item The model provider fully controls the model training process, including model structure design, hyperparameter tuning, optimization, and training schedule.
    \item The model provider only shares the weight of the trained model, but not the training data due to the scarcity and sensitivity of the critical domain \cite{kirkland_neuromorphic_2020} of the neuromorphic data. 
\end{itemize}

\subsubsection{Model consumer} 
The model consumer acquires pre-trained SNN models from the model providers and deploys the models for inference, often referred to as MLaaS~\cite{wu_graphguard_2024}.
\begin{itemize}[topsep=1pt]
    \item The model consumer is incapable of independently collecting sufficient labeled neuromorphic data for training.
    \item The model consumer has no access to neuromorphic hardware such as an Intel Loihi \cite{orchard_efficient_2021} or Von Neumann architecture computer with sufficient computational power to independently train an SNN model.
    \item The model consumer may be able to collect a small amount of unsellable data in the problem domain.
\end{itemize}

\subsection{Threat Model}
The paper considers the security risk of a backdoor attack in SNN models obtained from an untrustworthy model provider. Highlighting the more practical data-free assumption for backdoor detection and the label-free assumption for backdoor mitigation distinguishes our research from previous literature.
\subsubsection{Attacker's Goals and Capabilities Assumptions}

This paper considers the model provider as the attacker aiming to conduct a classical data poisoning-based, dirty-label all-to-one backdoor attack against the model consumer. The attacker is motivated to exploit the victim's system by triggering the compromised model to perform abnormally in a predefined way, such as bypassing the facial recognition security check \cite{li_light_2020}. The attack focuses on the goal of successfully conducting a backdoor attack while avoiding suspicion from the defender, that is, maximizing ASR and CA with a lower L2 norm of the attack pattern.

This research follows the classical assumption of the attacker's capabilities from previous studies \cite{gu_badnets_2019, doan_lira_2021, salem_dynamic_2022, abad_poster_2022, abad_sneaky_2024,liu_reflection_2020}.
\begin{itemize}[topsep=1pt]
    \item The attacker has full access to modify the model training process to suit the attack goal.
    \item The attacker has full access to freely modify the training data and the corresponding labels digitally to suit the attack goal.
    \item The attacker has sufficient knowledge and computational resources to perform the latest and most advanced backdoor attack strategies. Such as the SOTA dynamic trigger pattern attack \cite{abad_sneaky_2024} to maximize the effectiveness and stealthiness of the attack shown in Equation \eqref{eq:dynamic}. 
\end{itemize}

\subsubsection{Defender's Goals and Capabilities Assumptions}

This paper considers the model consumer to be the defenders motivated by maintaining the acquired model function as expected. The defender aims to perform a post-training model-based backdoor detection to identify a composed model containing a backdoor. The defender has the goal of detecting the backdoor attack and the corresponding ATC with high detection accuracy. The defender can use alternative models if the backdoor attack is detected and alternative models are available for the same problem domain. Otherwise, if replacement classifiers are not available, the attacker would want to mitigate the compromised model by suppressing the backdoor attack. Backdoor mitigation aims to reduce ASR while maintaining CA as high as possible.

This research follows a more strict and practical assumption on a less powerful defender than mainstream research on post-training defense. The motivation is to improve the robustness of the proposed defense so that it is also applicable to other relaxed scenarios.
\begin{itemize}[topsep=1pt]
    \item The defender has white-box access to the SNN model.
    \item The defender has no prior knowledge of the existence of the attack, the type of attack, or the attack target label. 
    \item The defender has no access to the training data, clean or poisoned, used to train the model.
    \item The defender has no access to clean reference models from the same domain; otherwise, they would deploy such model directly.
    \item Unlike the classical setting, the defender is incapable of collecting any data in the problem domain during the backdoor detection.
    \item The defender is capable of collecting a small set of unlabeled data in the same domain during backdoor mitigation.
\end{itemize}

\section{Temporal Membrane Potential Backdoor Detection}

In this section, we propose TMPBD, a data-free unsupervised backdoor detection framework. TMPBD detects if there is a backdoor attack embedded in the trained SNN model and its corresponding attack target label.  The TMPBD utilizes unsupervised hypothesis testing to identify abnormally high decision boundaries from the backdoor ATC to the benign classes.  We innovatively quantify the prediction confidence of SNNs with TMP and quantify the decision boundaries with MM. The MM is estimated by generating and optimizing the synthetic input that maximizes the MM. In this section, we present the design rationales, supported by both conceptual reasoning and empirical results, and demonstrate detailed implementation procedures.

\subsection{Design Intuition}
\label{design_intuition}
First, we show that the presence of a backdoor is associated with abnormal overfitting, manifested as inflated prediction confidence for the attack target class. To achieve the attack objective, a trigger (pattern) with a small spatiotemporal footprint is crafted to exert a disproportionately large influence, steering the model toward the target class. For the backdoor attack to succeed on triggered inputs, its effect must dominate the sample's genuine class-discriminative features so that a triggered sample is not classified as its true class. This “small fraction of data, large effect” phenomenon induces overfitting that can persistently bias the model, sometimes even visible on clean or random inputs.

\begin{figure}[t]
    \centering
    \includegraphics[width=\linewidth]{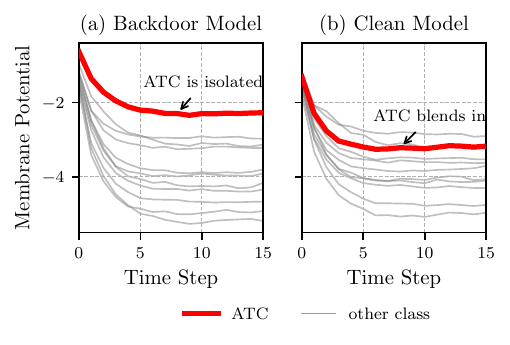}
    \caption{Membrane potential over time for the attack target class (ATC, red) in (a) backdoor and (b) clean models, averaged over the clean test set.}
    \label{fig:mpc}
\end{figure}

Figure~\ref{fig:mpc} illustrates this effect using membrane potential as a proxy for prediction confidence for class~0 (bold red curve). In the poisoned model, the red curve is markedly higher than the others, indicating a systematic bias toward the target class even when the inputs are trigger-free. In the clean model, the red curve is indistinguishable from the rest.

The example shown uses a static backdoor under the default configuration from prior work~\cite{abad_sneaky_2024} (see Appendix~\ref{app:backdoor}) on the DVS128-Gesture dataset, but the theorem is not tied to this attack type; the detection mechanism we derive from it remains effective across diverse backdoor variants in the experiments below.
Secondly, we justify the intuition behind using the TMP as the quantitative representation of the confidence of prediction. The TMP indicating $\frac{1}{T} \sum_{t=0}^{T} \hat{V}_{c,t}(\mathbf{x})$ is the average membrane potential over the timestamp of neurons from the last spiking layer. TMP also averages the membrane potential on all neurons corresponding to the same output label because of the common practice of adding a voting layer after the output spiking layer to increase robustness by having multiple output neurons correspond to the same output label vote for the final decision in SNNs \cite{fang_incorporating_2021}. The empirical studies in Figure \ref{fig:mpc} have shown the effectiveness of the membrane potential in capturing the confidence in prediction. TMP aims to concentrate the membrane potential series, the red trending line, into a single value.

\begin{figure}[t]
    \centering
    \includegraphics[width=\linewidth]{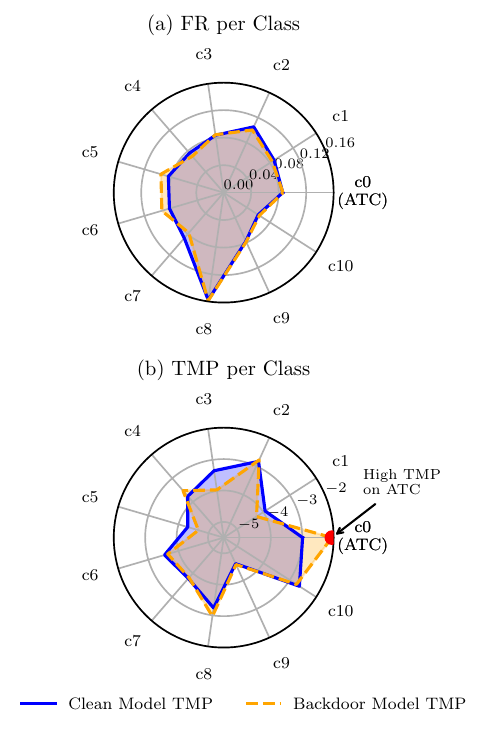}
    \caption{Class-wise (a) firing rate (FR) and (b) temporal membrane potential (TMP) between clean and backdoor models, averaged over the clean test set. Each axis corresponds to one of the 11 classes (c0--c10), where c0 is the attack target class (ATC).}

    \label{fig:radar}
\end{figure}

We argue that the proposed TMP is a more effective measurement of prediction confidence than commonly used alternatives, namely, firing rate (FR) and the highest membrane potential (HMP). 
In spiking neural networks (SNNs), FR is often considered analogous to the concept of logits in artificial neural networks (ANNs), where logits represent the pre-softmax activation values of the final layer. 
While logits are widely utilized in existing backdoor defense techniques, they do not naturally exist in SNNs due to their threshold-based nonlinearity. 
This makes FR a limited proxy for prediction confidence, as it merely reflects time-averaged spike counts rather than fine-grained membrane dynamics. 
To validate our hypothesis, we extend the empirical analysis and demonstrate that TMP more effectively captures backdoor-induced overfitting than FR, as illustrated in Figure~\ref{fig:radar}.
As shown in Figure~\ref{fig:radar}, the FR values averaged over the clean test set do not exhibit a notable difference between the clean and backdoor models. 
In contrast, TMP shows a pronounced gap at class c0, corresponding to the ATC. 
This suggests that while the backdoor fails to trigger sufficient spiking activity to alter FR, it still results in significantly elevated membrane potential toward the target class. 
This supports our claim that TMP provides a finer-grained and more sensitive view of the backdoor effect, revealing confidence shifts that FR fails to capture.

To demonstrate that TMP is a better design choice than HMP, we performed a small-scale backdoor detection experiment comparing backdoor detection behavior and accuracy between TMP and HMP on the Gesture-DVS benchmark dataset with static and dynamic attack patterns \cite{abad_sneaky_2024} and a clean control group. 
Table \ref{table:hmp_tmp} shows that the backdoor detection employing TMP exhibits better prediction accuracy compared to that of HMP. To be more precise, the box plot showing the distribution of p-value in backdoor detection with HMP and TMP is shown in Figure \ref{fig:radar}. 
We observe that having a higher p-value for the clean model and a lower value for poisoned models is better.  
Overall, the TMP-based algorithm is more accurate in prediction and more confident in such correct prediction, reflected as the distribution is farther away from the decision boundary at 0.05, denoted as the dotted red horizontal line.
\begin{table}[t]
\centering
\renewcommand{\arraystretch}{1.3}
\setlength{\tabcolsep}{6pt}  
\small
\begin{tabular}{c|ccc}
\hline
          & \textbf{Clean} & \textbf{Static} & \textbf{Dynamic} \\ \hline
HMP       & 80\%           & 90\%            & 100\%            \\ \hline
TMP       & 90\%           & 100\%           & 100\%            \\ \hline
\end{tabular}
\vspace{0.5em}
\caption{Backdoor detection accuracy (\%) of HMP and TMP on the DVS128-Gesture dataset.}
\label{table:hmp_tmp}
\end{table}
\begin{figure}[t]
    \centering
    \includegraphics[width=\linewidth]{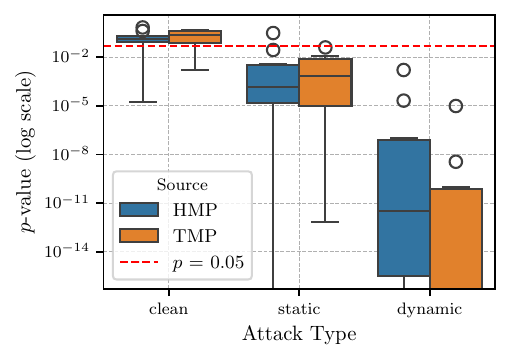}
    \caption{Boxplot of $p$-values for hypothesis testing under different attack types, using highest membrane potential (HMP) and temporal membrane potential (TMP) as prediction confidence measures.}
    \label{fig:tmp-hmp}
\end{figure}
Finally, we adopt MM statistics to characterize decision-making under synthetic stimuli. Because the TMP distribution differs across models and datasets, we detect anomalies in the MM of the TMP, not in the raw TMP, ensuring robustness to pseudo-sample bias and eliminating any need for real data. Here, we seek a synthetic input $x$ whose margin between the ATC and all other classes exceeds any margin among non-ATC pairs. In SNNs, this condition is formalized as:
\begin{multline}
    \max_{\mathbf{x} \in \mathcal{X}} \left[ \frac{1}{T} \sum_{t=0}^{T} \hat{V}_{a,t}(\mathbf{x}) - \max_{k \in \mathcal{Y} \setminus a} \frac{1}{T} \sum_{t=0}^{T} \hat{V}_{k,t}(\mathbf{x}) \right] \\ \gg \max_{\mathbf{x} \in \mathcal{X}} \left[ \frac{1}{T} \sum_{t=0}^{T} \hat{V}_{c,t}(\mathbf{x}) - \max_{k' \in \mathcal{Y} \setminus c} \frac{1}{T} \sum_{t=0}^{T} \hat{V}_{k',t}(\mathbf{x}) \right]
\end{multline}
Where $a\in\mathcal{Y}$ denotes the backdoor ATC and $c \in \mathcal{Y} \setminus a$ represents any of the benign labels. $\hat{V}_{a,t}(\mathbf{x})$ here denotes the membrane potential value at time $t$ in the neuron of the last spiking layer that corresponds to the output class $a$. This observation forms the backbone of our proposed backdoor detection strategies.

\subsection{Detection Procedure}
The procedure of the proposed unsupervised data-free backdoor detection method comprises two parts: the estimation stage and the detection stage.

\textbf{Estimation stage} aims to generate and optimize the neuromorphic samples input independently for each label to find and estimate the MM statistics for the TMP corresponding to each class $c \in \mathcal{Y}$. The MM statistics for TMP are denoted as $r_c$ and estimated by using gradient descent to solve:
\begin{equation}
    r_c = \max_{x \in \mathcal{X}} \left( \frac{1}{T} \sum_{t=0}^{T} \hat{V}_{c,t}(\mathbf{x}) - \max_{k \in \mathcal{Y} \setminus \{c\}} \frac{1}{T} \sum_{t=0}^{T} \hat{V}_{k,t}(\mathbf{x}) \right)
\end{equation}
The $x$ optimization via gradient ascent is guaranteed to converge smoothly to the local maximum in our experimental setting. Thus, we follow the common practice of optimizing multiple uniformly randomly initialized samples in parallel to estimate the global maximum with the largest local maximum. This guarantee was from Theorem 3.2 in \cite{bubeck_convex_2015} that the TMP is bounded and Lipschitz, since the input data $x$ is a closed convex set. This theorem has been thoroughly illustrated and proven in the ANNs realm with RGB image input \cite{MM-BD}, and we argue that the theorem holds true after porting it into the SNNs realm with neuromorphic data.

The DVS camera has a circuit time constant $\tau$ that describes the reaction time depending on the hardware, usually varying from 1-100 ms~\cite{chen_training_2023}. The time constant for which discretization continues stream events into instantaneous events. With this minimal gap between events, there is an upper bound for the number of events within a fixed total capturing time. The upper bound on the total number of events is carried over as the upper bound on the event count at each frame after integration into $T$ time frames. The count is nonnegative and bounded, so that the linear combination of frames falls in the same range, making it a closed convex set, the same as the RGB image. On the other hand, the membrane potential $\hat{V}_{c,t}(\mathbf{x})$ is bounded by $V_\text{threshold}$ and reflects the accumulation of discrete bounded inputs. Therefore, the TMP is also bounded and Lipschitz.

\textbf{Detection stage} conducts the anomaly detection framework proposed in the paper by Xiang et al. \cite{xiang_detection_2022} utilizing hypothesis testing based on Gamma distribution. The hypothesis test compares the chance that the largest MM in all classes $r_{\text{max}} = \max_{c \in \mathcal{Y}} r_c$ fits as the largest value of the distribution of the rest of the MM $r_{\text{rest}} = \{r_c \mid c \in y, r_c \neq r_{\text{max}}\}$. We have a hypothesis test with:
\begin{align*}
H_0 &: r_{\text{max}} \sim \text{Gamma}(r_{\text{rest}}), \text{no attack.} \\
H_a &: r_{\text{max}} \not\sim \text{Gamma}(r_{\text{rest}}), \text{attack exist.}
\end{align*}
Thus, we compute the order statistic p-value:
\begin{equation}
    \text{p-value} = 1 - H_0(r_{\text{max}})^{K}
\end{equation}
Here $H_0(r_{\text{max}})$ denotes the probability of $r_{\text{max}}$ belonging to the null distribution calculated from the Cumulative Distribution Function (CDF). Powered by $K$, the number of classes for order statistics due to $r_{\text{max}}$ was the maximum value among multiple statistics instead of individual statistics. 

The calculated p-value or false positive rate describes the chance of false rejection $H_0$. The equivalent of predicting the model is compromised when the model is actually clean. The p-value is then compared with the classical significance level $\alpha=0.05$. If the p-value $< 0.05$, the null hypothesis is rejected, suggesting that there is a backdoor attack in the model with an ATC associated with $r_\text{max}$. Otherwise, there is no attack.

\section{Neural Dendrites Suppression Backdoor Mitigation}

In this section, we propose NDSBM, a novel unsupervised backdoor mitigation technique for the scenario when the defender has no access to alternative models other than the detected compromised model. The method requires the defender to be capable of collecting a small amount of clean unlabeled data from the same problem domain. The proposed mitigation is based on the idea that the abnormally overfitted large TMP in the last spiking layer is accumulated from the slightly higher than normal output of neurons in the early layers associated with attack trigger patterns. By suppressing such effect, we can effectively "unlearn"~\cite{xia_edge_2025} the backdoor behavior embedded in the poisoned model.

\subsection{Design Intuition}

Although clamp-based mitigation has been explored in ANN \cite{MM-BD}. We still face a number of technical challenges. In ANN, the defender clamps on the activation value, which is impossible in SNN, as the neuron output is binary. The clamping on the membrane potential is also impossible, as it is already clamped by  $V_\text{reset}$ and  $V_\text{threshold}$. Therefore, we creatively clamp the input to the neurons. In SNN, the input of a neuron is equivalent to the weight of neural dendrites connecting neurons of two layers that control the decay of the neural signal. Mathematically, that is due to the only non-zero output from the previous layer being multiplied by the weight being one.

The activation value in ANN is guaranteed to be non-negative by the nature of the ReLU activation function \cite{fukushima_cognitron_1975}. In contrast, the weights in SNN can be negative. We chose to dually clamp with distinct floor and ceiling values. There exist alternative approaches: max clamping, which only has a ceiling, and absolute clamping, which floor has negative but the same magnitude as the ceiling. The empirical comparison between performance across different design choices is discussed in the main experiment in Table \ref{tab:mitigation}.

\subsection{Mitigation Procedure}
The proposed NDSBM introduced dual clamping layers after the first convolution layer of each convolution block to clamp the input value $X_t$ to neurons after the clamping layer. The values are clamped between the ceiling $\mathbf{C}$ and the floor $\mathbf{F}$ to mitigate the backdoor effect early on. This adds the additional clamping layer on top of the normal behavior of the LIF neuron from Equation \eqref{eq:1} to:
\begin{multline} \label{eq:7}
H_{\text{clamp},t}(\mathbf{C,F}) = \\ V_{t - 1} + \frac{1}{\tau} \left( max(\mathbf{F},min(\mathbf{C},X_t) - \left( V_{t - 1} - V_{\text{reset}} \right) \right)
\end{multline}
Please note that in SNNs, the output of spiking neurons $S_{j,t}$ can only take the value of 1 if a spike and 0 otherwise. This behavior is described in Equations \eqref{eq:2} and \ref{eq:3}. Therefore, clamping $X_{t,i}$ is clamping the weight \( W_{ij}\) describing the direction and strength of the neural dendrite, or the connection between neurons in contiguous layers. By narrowing the clamping range, mitigation is more likely to filter out abnormal weights that relate to the backdoor pattern, but also increases the chance of falsely clamping clean weights. Therefore, the small clean set is used to observe the behavior of the model and to find suitable clamping parameters $\mathbf{C,F}$ to balance the degradation in CA and reduce the amount of ASR. The parameters are obtained by optimizing the following loss function:
\begin{multline}
\mathcal{L}_{\text{base}}\bigl(\mathbf{C},\mathbf{F},\lambda;\mathcal D\bigr)=\\[2pt]
\frac{1}{|\mathcal D|\,|\mathcal Y|}\!
\sum_{(x,y)\in\mathcal D}\sum_{c\in\mathcal Y}
\biggl[
  \Bigl(
    \frac{1}{T}\sum_{t=0}^{T-1}\hat V_{\text{clamp},c,t}(\mathbf{C},\mathbf{F})
  \Bigr)^{\!2}
  -
  \Bigl(
    \frac{1}{T}\sum_{t=0}^{T-1}\hat V_{c,t}
  \Bigr)^{\!2}
\biggr] \\[4pt]
\;+\;
\lambda\sum_{l=1}^{L}\bigl(\lVert\mathbf c_l\rVert_2+\lVert\mathbf f_l\rVert_2\bigr).
\end{multline}

CA is maintained by ensuring the same distribution of TMP on clean samples. The clamp covers $L$ convolution blocks to mitigate backdoor patterns in both the pixel and feature space. The clamping parameters are sufficiently large, so no clamping takes effect at the beginning. They are gradually reduced and motivated by the L2 norm loss term to narrow down the clamping range to filter out abnormal weights for low ASR goals. The weight term $\lambda$ helps to balance the conflict in achieving both goals.

\subsection{Full-Life-Cycle-Defense}
NDSBM can be further strengthened when cooperating with TMPBD. The end-to-end strategy starts by detecting the potential target label of the attack. The method flags the samples that are predicted to the target label by the original classifier. The clamped model is applied only to the suspicious sample. The idea is to avoid unnecessary mitigation on trustworthy samples.

\section{Experiment}

To ensure the effectiveness of our proposed defense, we follow the optimal experimental setup suggested in the original SOTA attack paper \cite{abad_sneaky_2024}, including model architecture, training protocols, and attack configurations. In the optimal setting, all attacks tested in the experiment: static, moving \cite{abad_poster_2022}, and dynamic \cite{abad_sneaky_2024} backdoor attacks reach 100\% ASR with near-negligible CA degradation. Note that the smart trigger backdoor attack \cite{abad_sneaky_2024} is discarded from the experiment because it is not maintained by 100\% ASR on all datasets. We benchmark on the three widely recognized neuromorphic benchmark datasets, which are DVS128-Gesture \cite{amir_low_2017}, CIFAR10-DVS \cite{li_cifar10-dvs_2017}, and N-Caltech101 \cite{orchard_converting_2015}, for complete validation on generalizability and robustness of the method. The proposed algorithm is implemented on the Spikingjelly neuromorphic computing framework \cite{fang_spikingjelly_2023} known for its universality on both the Von Neumann architecture platform and the neuromorphic platforms \cite{orchard_efficient_2021}, boosting the practical relevance of the proposed defense.

For each combination of attack method and dataset, we repeat the experiments ten times to ensure the robustness of the results. Each run uses a different attack target label to ensure that the choice of target label does not influence the results. DVS128-Gesture data attacks on the first ten labels. CIFAR10-DVS data attacks on all ten labels. N-Caltech101 attack on randomly selected ten labels. With random seed 42, selected classes include: $[81, 14, 3, 94, 35, 31, 28, 17, 13, 86]$. The ten clean repeated models for the control group are trained with different random seeds from zero to nine. All other randomness processes are configured with the same classical random seed 42 to ensure the reproducibility of the experiment. 

For reproducibility, the complete source code for all implementations and experiments is publicly available at \url{https://github.com/alexjiachenli/TMPBD-NDSBM}.

\subsection{Backdoor Detection}
\begin{table*}[t]
\centering
\renewcommand{\arraystretch}{1.2}
\setlength{\tabcolsep}{5pt}
\small
\begin{tabular}{c|c|cccc|cccc|cccc}
\hline
\multirow{2}{*}{} & \multirow{2}{*}{N} & \multicolumn{4}{c|}{DVS128-Gesture} & \multicolumn{4}{c|}{CIFAR10-DVS} & \multicolumn{4}{c}{N-Caltech101} \\ \cline{3-14} 
& & Clean & Static & Moving & Dynamic & Clean & Static & Moving & Dynamic & Clean & Static & Moving & Dynamic \\ \hline
\multicolumn{14}{c}{Backdoor Detection Accuracy} \\ \hline
NC~\cite{wang_neural_2019}     & 50 & \textbf{100\%} & 0\% & 0\% & 0\% & 0\% & \textbf{100\%} & 0\% & \textbf{100\%} & 0\% & \textbf{100\%} & \textbf{100\%} & \textbf{100\%} \\ \hline
ABS~\cite{liu_abs_2019}    & 50 & \textbf{100\%} & 0\% & 0\% & 0\% & \textbf{100\%} & 0\% & 0\% & 0\% & \textbf{100\%} & 0\% & 0\% & 0\% \\ \hline
NS~\cite{liu_abs_2019,bailey_fine-pruning_2018}     & 0  & \textbf{100\%} & 10\% & 50\% & 50\% & 60\% & 90\% & 50\% & \textbf{100\%} & 60\% & 80\% & 0\% & 50\% \\ \hline
MMBD~\cite{MM-BD}   & 0  & \textbf{100\%} & 0\% & 80\% & 10\% & \textbf{100\%} & 0\% & 30\% & 0\% & 80\% & 20\% & 70\% & 50\% \\ \hline
TMPBD   & 0  & 90\% & \textbf{100\%} & \textbf{100\%} & \textbf{100\%} & 80\% & \textbf{100\%} & \textbf{100\%} & \textbf{100\%} & 90\% & 90\% & 90\% & \textbf{100\%} \\ \hline
\multicolumn{14}{c}{Attack Label Detection Accuracy} \\ \hline
NC~\cite{wang_neural_2019}     & 50 & \textbf{100\%} & 0\% & 0\% & 0\% & 0\% & 10\% & 0\% & 10\% & 0\% & 0\% & 0\% & 0\% \\ \hline
ABS~\cite{liu_abs_2019}    & 50 & \textbf{100\%} & 0\% & 0\% & 0\% & \textbf{100\%} & 0\% & 0\% & 0\% & \textbf{100\%} & 0\% & 0\% & 0\% \\ \hline
NS~\cite{liu_abs_2019,bailey_fine-pruning_2018}     & 0  & \textbf{100\%} & 0\% & 10\% & 0\% & 60\% & 0\% & 0\% & 20\% & 60\% & 0\% & 0\% & 0\% \\ \hline
MMBD~\cite{MM-BD}   & 0  & \textbf{100\%} & 0\% & 10\% & 0\% & \textbf{100\%} & 0\% & 10\% & 0\% & 80\% & 10\% & 0\% & 50\% \\ \hline
TMPBD  & 0  & 90\% & \textbf{100\%} & \textbf{100\%} & \textbf{100\%} & 80\% & \textbf{100\%} & \textbf{100\%} & \textbf{100\%} & 90\% & \textbf{90\%} & \textbf{70\%} & \textbf{100\%} \\ \hline
\end{tabular}
\vspace{0.5em}
\caption{
Backdoor and attack label detection accuracy (\%) of our proposed TMPBD against various defense methods across three neuromorphic datasets (DVS128-Gesture, CIFAR10-DVS, and N-Caltech101), evaluated under four attack types: Clean, Static trigger, Moving trigger, and Dynamic trigger. \textbf{N} denotes the number of samples per class used during detection. The highest accuracy in each row is highlighted in bold.
}
\label{tab:detection}
\end{table*}

In this section, we evaluate the backdoor detection accuracy and attack label detection accuracy of our proposed TMPBD compared to the ANN defense adopted as the baseline. Although theoretical deficiencies prevented existing ANN backdoor defenses from working effectively on SNNs have been explored. Due to a lack of detailed experimental results or open-sourced implementations. We re-implement the existing defense to empirically validate the deficiencies. The ANN backdoor detection methods adopted for SNN in this experiment are NC \cite{wang_neural_2019}, ABS \cite{liu_abs_2019}, Neuron Simulation (NS), and MMBD \cite{MM-BD}. Each defense is evaluated in 12 scenarios, combining three data sets with four model conditions (three attacks and one clean control group). Each scenario is repeated ten times to ensure robustness. The detection accuracy is calculated from the ratio of correct detections from 10 repetitions. The detection hyperparameters are chosen on the basis of the VRAM limit and the convergence speed of the data sets. Here we initialize 3 parallel synthetic inputs and optimize for 5000 epochs.  

To ensure that the experiment on our adoption of the ANN defense is reproducible, we introduce the defense configuration in the following list. The information is also available in our published source code. 
\begin{itemize}[topsep=1pt]
    \item In NC \cite{wang_neural_2019}, for each label $c \in \mathcal{Y}$, we uniformly randomize a learnable putative trigger in the same dimension as the data $x$ with a value ranging from 0 to 0.1. We regularize the trigger by the L1 norm during optimization. We used a Median Absolute Deviation (MAD) threshold of 2 for the anomaly detection, which is equivalent to the confidence level 95\% in the normal distribution. We use FR for computing binary cross-entropy in the loss function.
    \item We adopt ABS \cite{liu_abs_2019} differently from the previous literature \cite{abad_sneaky_2024}. For each neuron in the last layer, we generate and optimize a synthetic input to maximize the FR of the neuron. The synthetic input is generated in the same way as a putative trigger in NC. During optimization, additional clamping is implemented in the range $[0,1]$ to match the range of non-negative legal input. We calculate the average FR of each neuron after taking all synthetic inputs and mark the neuron suspicious if the average FR exceeds the threshold, which is 95\% percentile in our implementation. The ABS then generates a putative trigger that can maximize the FR of all suspicious neurons. Finally, the algorithm clips the putative trigger on the clean dataset and checks whether a variation in model prediction results from the trigger.
    \item NS is not an existing standalone defense, but the suspicious neutron detection part in other defense frameworks such as ABS~\cite{liu_abs_2019} or fine-pruning~\cite{bailey_fine-pruning_2018}. We consider it as a standalone defense, as it does not require access to clean samples. We interpret the non-empty output of the suspicious neuron list as a backdoor detected with attack labels associated with suspicious neurons.
    \item In MMBD \cite{MM-BD}, we replace logits with FR. We initialize three samples uniformly between 0 and 1 to optimize in parallel for a maximum of 5000 epochs to incorporate the additional memory and complexity of the neuromorphic data.
\end{itemize}
The results of the experiment are shown in Table \ref{tab:detection}. Our proposed defense outperforms all existing defenses in detecting the existence of backdoors and attack labels on compromised models and falls short only by a small margin in detecting clean models. Among the adopted defenses, the two backdoor pattern reverser engineer-based detections, NC and ABS, despite having additional access to a small clean sample, have failed catastrophically. The NC makes dataset-specific predictions independent of the attack type. The NC predicts no attack for all DVS128-Gesture data, attack target label 2 for all CIFAR10-DVS models, and attack on label 89 for all N-Caltech101 models. However, ABS has successfully identified numerous suspect neurons associated with output labels, but failed to validate the existence of backdoor attacks with the putative trigger pattern. As a result, ABS predicts that all models are clean. This indicates that the failure of the reverse engineer-based approach in SNNs is potentially due to the exponentially larger search space of neuromorphic data. 

By excision of the pattern, reverse engineering, and validation step, the NS shows acceptable backdoor detection accuracy, especially in two datasets transformed from the original static image form.  However, NS fails to locate the attack target label even after detecting the attack. The MMBD is worse at detecting backdoors compared to neuron simulation, but locates the target label more accurately once the attack is detected. The advantage is inferred from the robustness of the MM statistic over the absolute value of FR during optimization. The performance of those two defenses further validates the theorem that the backdoor causes overfitting, especially for the dynamically triggered attack samples. Our proposed method improves over MMBD by utilizing the MM statistic of TMP instead of FR, which is more informative.

\subsection{Backdoor Mitigation}
\begin{table*}[t]
\centering
\renewcommand{\arraystretch}{1.3}
\setlength{\tabcolsep}{4pt}
\small
\begin{tabular}{c|cc|cc|cc|cc}
\hline
\multirow{2}{*}{} & \multicolumn{2}{c|}{Clean} & \multicolumn{2}{c|}{Static} & \multicolumn{2}{c|}{Moving} & \multicolumn{2}{c}{Dynamic} \\ \cline{2-9}
& CA(\%) $\uparrow$ & ASR(\%) $\downarrow$ & CA(\%) $\uparrow$ & ASR(\%) $\downarrow$ & CA(\%) $\uparrow$ & ASR(\%) $\downarrow$ & CA(\%) $\uparrow$ & ASR(\%) $\downarrow$ \\ \hline
Original          & 97.65$\pm$1.03 & 0.31$\pm$0.99  & 98.09$\pm$0.99 & 100.00$\pm$0.00 & 97.21$\pm$1.29 & 100.00$\pm$0.00 & 84.71$\pm$12.48 & 100.00$\pm$0.00 \\ \hline
\multicolumn{9}{c}{Supervised mitigation methods requiring test labels} \\ \hline
Fine-Tuning~\cite{bailey_fine-pruning_2018}       & 64.56$\pm$6.63 & 3.00$\pm$4.70  & 56.32$\pm$6.97 & 4.38$\pm$11.26  & 70.29$\pm$13.98 & 5.91$\pm$12.98  & 88.53$\pm$5.54  & 3.28$\pm$4.51   \\ \hline
MMBM~\cite{MM-BD}     & 73.09$\pm$8.38 & 2.90$\pm$2.99  & 82.50$\pm$4.67 & 46.06$\pm$33.33 & 73.68$\pm$5.11  & 18.34$\pm$19.01 & 71.76$\pm$16.08 & 1.40$\pm$2.49   \\ \hline
\multicolumn{9}{c}{Unsupervised mitigation methods NOT requiring test labels} \\ \hline
Self-Tuning~\cite{bailey_fine-pruning_2018}       & 7.20$\pm$3.13  & 11.44$\pm$19.89 & 5.88$\pm$1.39  & 9.06$\pm$20.29  & 7.20$\pm$2.81   & 15.72$\pm$19.66 & 6.47$\pm$1.73   & 25.56$\pm$20.18 \\ \hline
Max Cla.          & 83.09$\pm$3.81 & 2.28$\pm$4.30  & 84.41$\pm$4.91 & 85.81$\pm$20.80 & 89.27$\pm$4.50  & 75.81$\pm$25.78 & 88.83$\pm$4.00  & 19.38$\pm$13.39 \\ \hline
Abs. Cla.         & 84.26$\pm$4.75 & 1.34$\pm$2.09  & 83.82$\pm$7.47 & 84.22$\pm$16.31 & 87.21$\pm$6.05  & 68.13$\pm$26.15 & 89.12$\pm$2.52  & 20.81$\pm$14.50 \\ \hline
NDSBM             & 72.50$\pm$6.43 & 3.69$\pm$10.27 & 72.21$\pm$6.56 & 30.41$\pm$25.92 & 83.38$\pm$8.29  & 29.87$\pm$19.92 & 89.86$\pm$3.21  & 8.44$\pm$9.91   \\ \hline
TMPBD+NDSBM       & 97.06$\pm$1.55 & 0.31$\pm$0.99 & 96.33$\pm$3.12 & 19.94$\pm$26.48 & 95.88$\pm$3.00  & 38.12$\pm$35.44 & 92.06$\pm$4.29  & 2.81$\pm$3.95   \\ \hline
\end{tabular}
\vspace{0.5em}
\caption{The average value and standard deviation of CA and ASR before and after mitigation with our proposed NDSBM and TMPBD+NDSBM frameworks against different backdoor attack mitigation defenses on the DVS128-Gesture dataset under clean, static, moving, and dynamic triggers.}
\label{tab:mitigation}
\end{table*}
The backdoor mitigation defense aims to alleviate the effect of the backdoor so that the poisoned model is no longer sensitive to trigger patterns. In this experiment, we focus on the DVS128-Dataset \cite{amir_low_2017} with the corresponding SNN model architecture \cite{fang_incorporating_2021} to compare the performance of different mitigation strategies on different trigger types in a controlled environment. The attack types involved in this experiment are a clean control group, static trigger attack, moving trigger attack \cite{abad_poster_2022}, and dynamic trigger attacks \cite{abad_sneaky_2024}.

The mitigation defense is evaluated by the ability to reduce ASR while maintaining CA. In this experiment, ASR and CA are evaluated by the test set, excluding the data involved in mitigation and training to avoid optimistic bias and lack of generalization. Furthermore, when assessing the ASR, the test sample with the same label as the target attack label is excluded following common practice \cite{karim_fisher_2024} due to the inability to identify the label because of the backdoor effect or discriminative characteristics of the class. The mathematical representation of ASR is shown below:
\begin{equation}
\label{eq:asr_exp}
    \text{ASR} = \frac{\left| \{ (x_i + \delta_i, \tilde{y}) \mid i \leq r, \, y_i \neq \tilde{y} \, \wedge \, h(x_i + \delta_i) = \tilde{y} \} \right|}{\left| \{ (x_i + \delta_i, \tilde{y}) \mid i \leq r, \, y_i \neq \tilde{y} \} \right|}
\end{equation}
We adopt supervised unlearning-based fine-tuning defense and clamping-based MMBM for SNNs for reference, as they would require access to the label information of the small clean set that violated our assumption on the defender's capability in the threat model. The necessary modifications to MMBM have been made to accommodate the SNN situation. Specifically, we set the weight amendment factor $a=1.2$, the learning rate to 0.1, the target CA to 85\%, and the initial $c=1e-5$. As discussed, due to the lack of activation and the threshold nature of membrane potential in SNNs, the MMBM are modified to max clamping the same weights as clamped by the proposed mitigation. The training accuracy is calculated via MSE over FR and one-hot encoded true label.

For the baseline, we modified the fine-tuning to use the predicted label as a putative label for fine-tuning, referred to as self-tuning. We also performed ablation tests on different clamping methods of the proposed weight clamping approach: max clamping, absolute clamping, and dual clamping. Finally, we experiment with the combination of the proposed attack label detection and dually clamping end-to-end backdoor defense pipeline. Note that although NC can mitigate after detection, the poor detection performance indicates that there is no experimental value for further mitigation with NC. Therefore, NC is discarded in the mitigation experiment. All of the defense runs for 50 epochs and have access to the first 20 samples from each class (around 2/3 of the total testing set).

The experiment is repeated ten times for each combination with a different attack target label or random seed for clean models. The average and standard deviation of CA and ASR of different mitigation methods in different types of attacks are shown in Table \ref{tab:mitigation}.

For supervised mitigation,  MMBM and fine-tuning outperform each other under different conditions with compromise, where lowering ASR often lowers CA simultaneously. In general, modified MMBM is more robust among supervised mitigation methods, especially since original MMBM is more computationally expensive to bypass \cite{MM-BD} compared to fine-tuning \cite{abad_sneaky_2024,riano_flashy_2024}. The observation demonstrates the feasibility of weight clamping. In a more practical but challenging unsupervised mitigation setting, self-turning fails due to a disastrous reduction in CA. The result shows that even with an original CA as high as 97.65\% on average, the accumulated putative label's small error destroyed the classifier's original behavior. Among all clamping approaches, dual clamping is most effective in reducing ASR, although it has a slight degradation in CA as a trade-off.

The main drawback of the proposed method is the non-negligible drop in CA. However, there is a workaround that only feeds suspicious samples to the clamped model. The suspicious samples are defined as samples predicted to the known attack target label by the compromised model. This solution is based on the accurate prediction of attack target labels, which has been achieved with our proposed TMPBD method, with an accuracy of 100\%. By combining the TMPBD with NDSBM, the pipeline can nearly completely eliminate backdoors in dynamic attacks and significantly reduce ASR in other attacks with negligible degradation in CA.

\section{Related Work}
\label{related_work}

Backdoor defense research for SNNs is still in its infancy. Most prior attempts simply port post-training defenses from ANNs, yet the spike-driven computation, binary activations, and temporally coded inputs of SNNs undermine their effectiveness.  We organize the literature by the defense mechanism and highlight, for each category, the SNN-specific obstacles that remain unsolved until our work.

\subsection{Activation-Analysis Defenses}

\textbf{Artificial Brain Stimulation (ABS).}  
ABS identifies neurons strongly correlated with an attacker’s target label, reconstructs a trigger, and tests it on clean inputs \cite{liu_abs_2019}.  When frames are collapsed for SNNs, the lack of ReLU “turn points” yields many false positives \cite{abad_sneaky_2024}.

\textbf{Maximum-Margin Backdoor Detection (MMBD).}  
MMBD replaces ABS’s turn-point heuristic with MM statistics of logits \cite{MM-BD}. Although this removes the ReLU dependency, the real-valued activations relied on by MMBD are absent in SNNs, so detection accuracy degrades in SNNs.

\subsection{Reverse-Engineering Defenses}

\textbf{Neural Cleanse (NC).}  
NC searches for a minimal per-class trigger and flags classes whose trigger is unusually small \cite{wang_neural_2019}.  In SNNs, the search space explodes (neuromorphic inputs have an extra temporal dimension) and the binary spike output provides little gradient guidance, so NC becomes prohibitively slow and inaccurate.

\textbf{Unsupervised Anomaly Detection (UAD).}  
Xiang et al.\ learn class-specific perturbations without training data and apply statistical testing on the perturbation norms \cite{xiang_detection_2020}.  Their optimization assumes softmax logits and has not been adapted to spike trains.

\subsection{Parameter-Repair Defenses}

\textbf{Fine-Pruning.}  
Removing low-contribution neurons can excise backdoor-related units in ANNs \cite{bailey_fine-pruning_2018}.  Because SNNs encode information in precise spike timings, nearly every neuron is indispensable; pruning slashes CA while barely reducing ASR \cite{abad_sneaky_2024}.

\textbf{Maximum-Margin Backdoor Mitigation (MMBM).}  
MMBM bounds suspicious activations during fine-tuning instead of deleting neurons \cite{MM-BD}.  Porting the method to SNNs requires clamping membrane potentials, which have not been systematically studied.

\subsection{Our Position in the Field}

The above defenses either \emph{(i)} depend on ReLU-style activations, \emph{(ii)} perform gradient-heavy trigger search infeasible for spike data, or \emph{(iii)} degrade CA because they treat SNN neurons like ANN activations.  
Our proposed TMPBD + NDSBM is the first full-lifecycle defense designed expressly for SNNs and overcomes each blocker:


\section{Threats to validity}
This section discusses the potential threats to the validity of our experimental results and how we address or mitigate them.

\subsection{Scalability}
One trade-off of any data-free backdoor detection method is the additional computational overhead incurred by the generation and optimization of synthetic data used for detection. However, we argue that scalability is not a blocker of the proposed TMPBD. The backdoor detection of a classifier for the DVS128-Gesture dataset is faster than model training (14 min 30 s vs. 18 min 45 s on RTX5090), while taking much less VRAM. For backdoor mitigation, NDSBM takes only 2 min 24 s to mitigate the discussed model. Moreover, TMPBD was cleverly designed to compute the MM of each class independently, suggesting that detection can be sped up by parallelization up to the factor of the number of classes, i.e., 11 for the DVS128-Gesture dataset.

\subsection{Adaptive Attacker}
\label{adaptive_attacker}
Here, we evaluate the robustness of TMPBD against adaptive attackers with defense knowledge. As discussed in Section \ref{design_intuition}, TMPBD detects abnormal overfitting phenomena resulting from backdoor attacks. Intuitively, an adaptive attacker would attempt to suppress such a phenomenon to bypass TMPBD. We consider two adaptive attack approaches: amplitude-suppression adaptation (ASA) and peak-alignment adaptation (PAA). ASA is designed to depress the absolute membrane potential of the ATC. PAA, inspired by adaptive attack from \cite{MM-BD}, attempts to align TMP of ATC with the largest non-target TMP by minimizing the margin. In the context of Figure~\ref{fig:mpc}, ASA blends the red line into the gray lines, while PAA squeezes the red line closer to the cluster of the gray lines.

Both approaches were achieved by introducing an additional term controlled by a loss-penalty weight to the loss function during the model training stage as follows:
\begin{align}
\mathcal{L}_{\text{ASA}}
  &= \mathcal{L}_{\text{MSE}}
     + \lambda_{\text{ASA}}
       \;\mathbb{E}_{x\sim D}\!\Bigl[
         \Bigl|\,
           \underbrace{%
             \frac{1}{T}\sum_{t=0}^{T-1}\hat V_{a,t}(x)
           }_{\text{ATC TMP}}
         \Bigr|
       \Bigr] 
\end{align}
\begin{equation}
\begin{aligned}
\mathcal{L}_{\text{PAA}}
  &= \mathcal{L}_{\text{MSE}}
     + \lambda_{\text{PAA}}
       \,\mathbb{E}_{x\sim D}\!\Bigl[
       \max\Bigl(
           0,\;
         \\ &   \underbrace{\tfrac1T\sum_{t=0}^{T-1}\hat V_{a,t}(x)}_{\text{ATC TMP}}
           -\!
           \underbrace{
             \max_{k\neq a}\;
             \tfrac1T\sum_{t=0}^{T-1}\hat V_{k,t}(x)
           }_{\text{largest non-target TMP}}
         \Bigr)
       \Bigr]
\end{aligned}
\end{equation}

\begin{figure}[t]
    \centering
    \includegraphics[width=\linewidth]{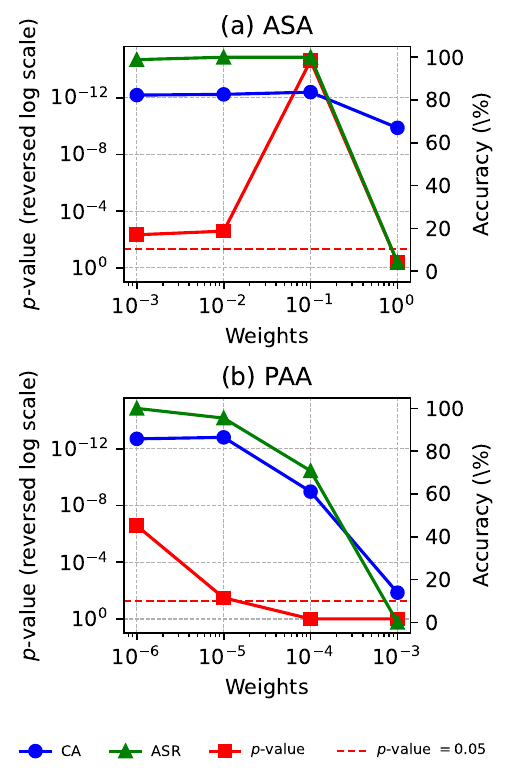}
    \caption{Statistical significance ($p$-value, left $y$-axis) and performance (CA, ASR, right $y$-axis) of TMPBD under (a) amplitude-suppression adaptation (ASA) and (b) peak-alignment adaptation (PAA) on the DVS128--Gesture dataset.}

    \label{fig:adaptive_attacks}
\end{figure}

We conduct the experiment on adaptive attack under the same setting as empirical studies from Section~\ref{design_intuition}, with a static attack on the DVS128-Gesture dataset. From the results shown in Figure~\ref{fig:adaptive_attacks}, we observe that the detection cannot be evaded unless we increase the penalty weight over a threshold, which would effectively also drop the CA and ASR to an impractically low level. Specifically, CA and ASR drop by 17.37\% and 95.83\% under ASA, 27.08\% and 28.79\% under PAA to bypass the detection. Such drastic performance degradation renders the backdoor practically useless, confirming TMPBD’s robustness against these adaptive strategies. 


Existing adaptive attack studies increase every non-target logit (equivalently, every non-target TMP in SNNs) to reduce the ATC margin, such as~\cite{MM-BD}. However, this approach introduces significant computation overheads, making it impractical. While existing adaptive attacks mainly focus on margin reduction, to our best knowledge, the distributed or timing-based evasions in SNNs have yet to be explored.

\subsection{Imbalanced Dataset}
\begin{figure}[t]
    \centering
    \includegraphics[width=\linewidth]{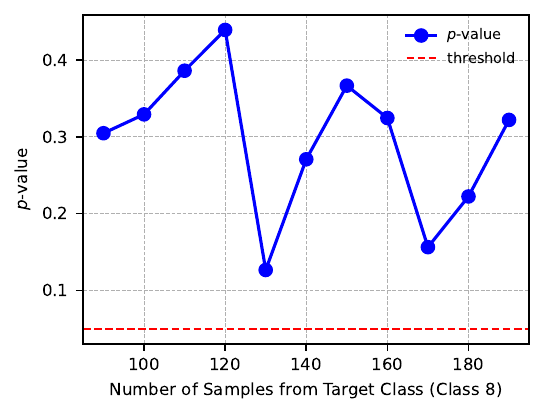}
    \caption{Detection $p$-values of clean models with varying sample counts from the target class (class 8) in the DVS128--Gesture dataset, with 90 samples fixed for other classes.}

    \label{fig:imbalance}
\end{figure}

The majority of model-based backdoor detection algorithms detect the overfitting of a small-sized backdoor pattern biased the model toward the ATC. Apart from backdoor attacks, imbalanced training data is another cause of overfitting, resulting in a model biased toward the majority class. Past literature suggests that MMBD falsely detects the clean model trained on severely imbalanced data as poisoned \cite{MM-BD}. However, the experiment in Figure \ref{fig:imbalance} shows that the detection results in TMPBD are invariant as the imbalance level increases. Training on more imbalanced data with more than double the sample of a class over others is impractical, as it harms the CA dramatically. 
\subsection{False Positive Issue}
Although the proposed TMPBD outperforms all existing defenses in the detection model with a backdoor. Detection is still occasionally too sensitive, and the clean model was misclassified as poisoned under the current default significant threshold $\alpha=0.05$. The flaw can be resolved with additional domain knowledge for domain-specific threshold tuning. This additional information is accessible under the classical threat model from the past literature. The results of significant threshold tuning are shown in Table \ref{tab:fpr}. Notably, for the CIFAR10-DVS data, reducing $\alpha$ to 0.02 would lower the FPR at no cost in the TPR, demonstrating the effectiveness of the adjustment.

\begin{table}[t]
\centering
\renewcommand{\arraystretch}{1.3}
\setlength{\tabcolsep}{6pt}
\small
\begin{tabular}{c|c|cccc}
\hline
\textbf{Dataset}  & $\alpha$ & \textbf{0.05} & \textbf{0.02} & \textbf{0.01} & \textbf{0.005} \\ \hline
\multirow{2}{*}{DVS128-Gesture} 
& TPR\% $\uparrow$ & 100 & 95  & 90  & 85  \\ \cline{2-6} 
& FPR\% $\downarrow$ & 20  & 20  & 10  & 10  \\ \hline
\multirow{2}{*}{CIFAR10-DVS}    
& TPR\% $\uparrow$ & 100 & 100 & 100 & 95  \\ \cline{2-6} 
& FPR\% $\downarrow$ & 20  & 10  & 10  & 10  \\ \hline
\multirow{2}{*}{N-Caltech101}   
& TPR\% $\uparrow$ & 95  & 95  & 90  & 90  \\ \cline{2-6} 
& FPR\% $\downarrow$ & 10  & 10  & 10  & 0   \\ \hline
\end{tabular}
\vspace{0.5em}
\caption{
True Positive Rate (TPR) and False Positive Rate (FPR) across datasets for different significance thresholds $\alpha$. The defender aims to maximize TPR while minimizing FPR.
}
\label{tab:fpr}
\end{table}

\subsection{All-to-All attacks}
\begin{figure}[t]
    \centering
    \includegraphics[width=\linewidth]{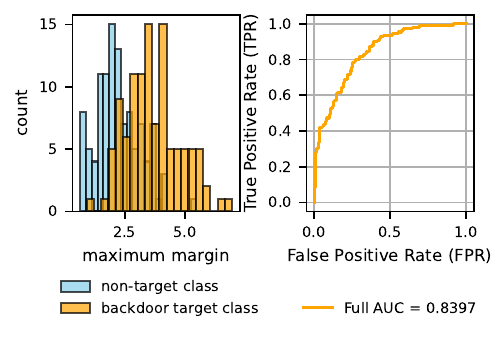}
    \caption{Analysis of maximum margin statistics and detection performance. (a) Margin distributions of clean classes (non-target) vs. ATC. (b) ROC curve of TMPBD detection, achieving full AUC = 0.8397 on DVS128-Gesture under all-to-all static attacks.}
    \label{fig:ata}
\end{figure}

The anomaly detection mechanism only validates the most suspicious class in TMPBD to detect all-to-one attacks. However, we observe a phenomenon, shown in Figure \ref{fig:ata}, similar to the previous literature \cite{MM-BD},  suggesting that the distribution of TMP MM of poisoned samples is distributed differently from clean samples with slight overlap, which serves as the basis for detection. The distribution was collected from ten clean models and ten all-to-all static attack models on DVS128-Gesture data. The attack pairs each neighboring class into the source class-ATC pair (that is, the sample with source class 0 is poisoned to class 1 with static trigger) \cite{gu_badnets_2019}. We argue that with access to additional domain knowledge that is accessible in a common defense setting \cite{liu_abs_2019}, we can calibrate the detection threshold to detect an arbitrary number of target classes of the backdoor in an attack of all-to-all or all-to-x without knowing the number of attack classes. Shown in Figure \ref{fig:ata}, the ROC curve of the margin-based detector with a full AUC of 0.8397, which shows a strong overall discriminative power. The result is particularly impressive given that this all-to-all attack is not effective and only has ASR 63.86$\pm$15.46\%. To our best knowledge, there are no proposed all-to-all attacks for SNN and we tried our best to adopt the current static \cite{abad_poster_2022} all-to-one attack for the all-to-all setting. Note that doubling the poison rate will counterintuitively not improve the effectiveness of the attack and reduce the ASR to 11.81\% and the dynamic attack \cite{abad_sneaky_2024} failed in the all-to-all setting with 0\% ASR. 
In particular, the proposed NDSBM backdoor mitigation method does not assume that the number of target classes remaining effective in the all-to-all attack reduced the ASR from 63.86$\pm$15.46\% to 16.04$\pm$7.36\%.

\subsection{Intrinsic Backdoored Data}
It has been an open problem to detect a backdoor attack for a model trained with intrinsic backdoored data \cite{MM-BD} for both SNN and ANN. The intrinsic backdoored data describes the data with class discriminative features that behave similarly to the backdoor pattern. For example, the model designed to classify if there is a backdoor in the sample always predicts true in the present backdoor pattern, which behaves as a backdoor-attacked model and will be detected as attacked. However, the model is, in fact, as designed to be, and is not under any backdoor attack. The phenomenon can also be observed in overly simple classification problems where the class discriminative features are as small as a few pixels. An example is N-MNIST \cite{orchard_converting_2015}, which uses a classifier that can make a high-confidence prediction based on a few pixels. 

\begin{figure}[t]
    \centering
    \includegraphics[width=\linewidth]{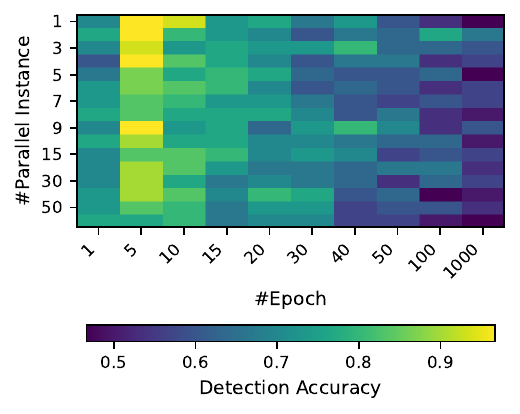}
    \caption{Heatmap of detection accuracy under grid search for different numbers of parallel instances and numbers of epoch with TMPBD for N-MNIST dataset.}
    \label{fig:heatmap}
\end{figure}

TMPBD is incapable of distinguishing an intrinsic backdoor from a real backdoor. However, by accessing additional domain knowledge, we can optimize an optimal detection hyperparameter to make TMPBD effective on such a dataset. Figure \ref{fig:heatmap} suggests that optimizing the hyperparameters of n-epoch and n-parallel on N-MNIST can lead to detection accuracy of up to 97\% via a "few-shot detection". We hypothesize that the intrinsic backdoor can be distinguished by the rate of convergence of the MM instead of the final optimized MM value. We hope that our first indicative finding toward solving this open question can inspire future research. 
\section{Conclusion}
This study addresses the challenge of backdoor attacks in Spiking Neural Networks by proposing two novel defenses. Temporal Membrane Potential Backdoor Detection (TMPBD) leverages the Maximum Margin statistic of temporal membrane potential to achieve unsupervised, post-training backdoor detection without the requirement of attack knowledge or additional data. Neural Dendrites Suppression Backdoor Mitigation (NDSBM) effectively reduces the backdoor effect while preserving clean accuracy through dual clamping of neural dendrites. The evaluations of the benchmark data sets demonstrated the near-optimal detection accuracy of TMPBD and the ability of NDSBM to lower the attack success rates to as low as 2.81\% on average with the help of TMPBD. The proposed defenses have outperformed all the existing backdoor defense techniques. The paper discussed the scalability and robustness of the proposed model under an adaptive attacker. The paper also illustrated that under a relaxed setting with access to domain knowledge, the proposed approach can be robust to imbalanced datasets, false positive issues, all-to-all attacks, and intrinsic backdoored data with minimal modification.

\section*{Acknowledgment}
This research has been supported by ARC Discovery Projects  DP190102835, DP220102803, DP240102140 and Linkage Project LP220200649. We thank the anonymous reviewers for their insightful suggestions
and comments.

\bibliographystyle{IEEEtran}
\bibliography{references}

\appendix

\section*{Experiment Detail}

\subsection{Details of Dataset}
\label{app:dataset}
The experiment is carried out in the widely used neuromorphic benchmark datasets DVS128-Gesture \cite{amir_low_2017}, CIFAR10-DVS \cite{li_cifar10-dvs_2017}, and N-Caltech101 \cite{orchard_converting_2015}. The techniques involved in backdoor attacks have been tested and have performed well in the original paper proposing the attacks \cite{abad_sneaky_2024}. The benchmark datasets are sufficiently complex for the classification task that practically represents the real-world scenario. They also cover a wide spectrum of data properties. The DVS128-Gesture dataset consists of human gesture movements with 29 different subjects under three different illumination conditions, directly captured by a DVS camera that is closest to the real-world situation. In contrast, the other two are pre-existing popular static images in the computer vision research field converted into neuromorphic data format via capture of the image showing on an LCD display with a DVS camera performing Repeated Closed-Loop Smooth (RCLS) Movement \cite{li_cifar10-dvs_2017}. Although the converted dataset is less practical, the CIFAR10-DVS provides the possibility of performance comparison with existing research in the field of ANNs, while the N-Caltech101 dataset contains 100 object classes plus one background class, offering insight into the situation with a large number of classification labels. 

Following the optimal setting from the reference paper \cite{abad_sneaky_2024}. We employ the same training settings as the original paper. Notable is the original paper choosing learning epochs of 28 for CIFAR10-DVS, 63 for DVS128-Gesture, and 30 for N-Caltech101 to align with the same CA in SOTA research \cite{samadzadehConvolutionalSpikingNeural2023}.

\subsection{Details of Training Configurations}
\label{app:training}
We adopted the commonly used corresponding network architecture for the classifier to defend in the related works \cite{abad_sneaky_2024,fang_incorporating_2021}. For the CIFAR10-DVS dataset, the network architecture comprises two convolutional layers, each followed by batch normalization and max pooling. This is succeeded by two fully connected layers with dropout, and a final voting layer of size ten to enhance classification robustness [17].
In contrast, the networks used for the DVS128-Gesture and N-Caltech101 datasets consist of five convolutional layers (with batch normalization and max pooling), two fully connected layers with dropout, and a voting layer.
Further architectural details are available in our code repository.

\subsection{Details of Backdoor Pattern}
\label{app:backdoor}
We follow the same recommended hypermeter for all attacks in the original literature \cite{abad_sneaky_2024} for all datasets to ensure a controlled environment and ensure that all attacks in all datasets are effective and reach 100\% ASR with nearly negligible CA degradation. Specifically, for static triggers, the trigger patch is located in the top left corner with a size of 10\% image with polarity=1. The attack pattern is static for all samples in all time steps. The pattern is injected into 10\% of the training set. The moving triggers are initialized similarly, but move two pixels to the right every time step. For the dynamic trigger, we set a hyperparameter $\alpha$, weight controlling the trade-off between CA and ASR in the loss function, to 0.5 to evenly balance CA and ASR. The visibility factor $\gamma$ is set to 0.01 to maximize stealthiness while maintaining high ASR and CA. 

\end{document}